\documentclass[useAMS]{mn2e}
\usepackage{times}
\usepackage{rotating}
\long\def\symbolfootnote[#1]#2{\begingroup%
\def\thefootnote{\fnsymbol{footnote}}\footnote[#1]{#2}\endgroup}
\newcommand{\gae}{\lower 2pt \hbox{$\, \buildrel {\scriptstyle >}\over {\scriptstyle
\sim}\,$}}
\newcommand{\lae}{\lower 2pt \hbox{$\, \buildrel {\scriptstyle <}\over {\scriptstyle
\sim}\,$}}
\newcommand{\aprop}{\lower 2pt \hbox{$\, \buildrel {\scriptstyle \propto}\over 
   {\scriptstyle \sim}\,$}}
 
\begin{document}

\title[A model for the multiwavelength data for TDE]
{A model for the multiwavelength radiation from tidal disruption event Swift 
J1644+57}

\author[Kumar, Barniol Duran, Bo\v{s}njak, Piran]{P. Kumar$^{1}$\thanks
{E-mail: pk@astro.as.utexas.edu, rbarniol@phys.huji.ac.il, zeljka.bosnjak@cea.fr, tsvi.piran@mail.huji.ac.il}, R. Barniol Duran$^{2}$\footnotemark[1], \v{Z}. Bo\v{s}njak$^{3}$\footnotemark[1] and
T. Piran$^{2}$\footnotemark[1] \\
$^{1}$Department of Astronomy, University of Texas at Austin, Austin, TX 78712, USA\\
$^{2}$Racah Institute for Physics, The Hebrew University, Jerusalem, 91904, Israel \\
$^{3}$Department of Physics, University of Rijeka, 51000 Rijeka, Croatia}

\date{Accepted ;
      Received ;
      in original form February 15, 2013}

\pagerange{\pageref{000}--\pageref{000}} \pubyear{2013}

\maketitle

\begin{abstract}

Gamma-ray observations of a stellar tidal disruption event (TDE) detected by 
the Swift satellite and follow up observations in radio, mm, infrared and 
x-ray bands have provided a rich data set to study accretion onto massive 
black holes, production of relativistic jets and their interaction with the 
surrounding medium. The radio and x-ray data for TDE Swift J1644+57
provide a conflicting picture regarding the energy in relativistic jet 
produced in this event: x-ray data suggest jet energy declining with time
as t$^{-5/3}$ whereas the nearly flat light curves in radio and mm bands 
lasting for about 100 days have been interpreted as evidence 
for the total energy output increasing with time. We show in this work 
that flat light curves do not require addition of energy to decelerating 
external shock (which produced radio and mm emission via synchrotron process), 
instead the flat behavior is due to inverse-Compton cooling of electrons 
by x-ray photons streaming through the external shock; the higher x-ray flux 
at earlier times cools electrons more efficiently thereby reducing the emergent 
synchrotron flux, and this effect weakens as the x-ray flux declines with time.

\end{abstract}


\section{Introduction}

When a star wanders too close to a massive black hole it is shredded
by the strong tidal gravity of the BH. In this process a fraction of 
the star is captured by the BH and is eventually accreted, and roughly half 
of the stellar mass is flung out on hyperbolic orbits
(Lacy et al. 1982; Rees 1988; Evans \& Kochanek 1989; Goodman \& Lee 1989; 
Ayal, Livio \& Piran 2000). Such an encounter is expected to occur at a 
rate of $10^{-3}$---$10^{-5}$ yr$^{-1}$ per $L_*$ galaxy (Magorrian \&
Tremaine 1999; Wang \& Merritt, 2004; Bower, 2011).
The accretion disk is expected to produce blackbody radiation in UV and 
soft x-ray bands from the region close to the BH, and a bright optical 
flash is produced by a super-Eddington outflow and by the irradiation and 
photo-ionization of unbound stellar debris (Strubbe \& Quataert 2009, 2011).

Generation of a relativistic jet is also expected to accompany a tidal
disruption event (TDE) as the
stellar material bound to the BH gravity is accreted over a period of time;
we know empirically that accreting BHs in active galactic nuclei (AGNs) 
produce jets moving at speed close to that of light with Lorentz factor
of order 10.
A number of theoretical ideas regarding jet generation and energy production
in the context of TDEs have been explored in several recent papers e.g., 
De Colle et al. 2012, Liu, Pe'er \& Loeb 2012, Tchekhovskoy et al. 2013).
The jet launched in a TDE interacts with the circum-nuclear 
medium (CNM) and produces synchrotron radiation over a broad frequency band 
from radio to infrared which Giannios \& Metzger (2011) predicted should be 
observable for several years (also Metzger et al. 2012).

The NASA Swift satellite discovered a transient event, Swift J1644+57, in
$\gamma$ \& x-ray bands two years ago (Burrows et al. 2011, Bloom et al. 2011)
 which was subsequently observed in radio, mm, 
infrared and x-ray bands for a time period of more than year
with good temporal coverage, e.g. Zauderer et al. (2011), Levan et al. (2011),
Berger et al. (2012), Saxton et al. (2012), Zauderer et al. (2013), and the data
is broadly in line with expectations of a TDE, e.g. Burrows et al. (2011).

Two other transient events observed in the last year have also been 
suggested to be TDE (Cenko et al. 2011, Gezari et al. 2012). However, 
the follow up observations for these events have been sparse, and so 
we consider in this paper only Swift J1644+57 and the puzzle posed by 
the rich data set for this event.

The mm \& radio light curves for Sw J1644+57 were found to be
flat/rising for $\sim10^2$days by Berger et al. (2012). These
authors suggested that this implies continuous injection of energy to 
the decelerating blast wave, which produced this radiation, for $\sim10^2$d. 
However, that requires about 20 times more energy in the blast 
wave\footnote{Using simple relativistic equipartition arguments 
Barniol Duran \& Piran (2013) have shown that in the context of a
standard synchrotron scenario, without IC cooling of electrons,
the requirement that the blast wave energy
should increase with time during the first $\sim 200$ days in order
to explain the radio light curve behavior, does not depend on the choice 
of the geometry of the jet, nor does it depend on the
precise spectral fitting of the radio data.} than the initial jet
energy that produced the strong X-ray signal which was of order 
$10^{52}$erg. This energy requirement of more than a few times
$10^{53}$erg is very difficult to satisfy for a TDE model. We show in 
this paper that the multi-wavelength light curves of Sw J1644+57 can be 
understood without requiring energy in the blast wave to increase with time.

In the next section we describe the main features of the data for
Swift J1644+57, and the main puzzles. We then describe in \S3 the
main idea of this paper that offers a simple solution to the puzzling 
data, and the application to Sw J1644+57 is provided in \S4. Some 
concluding remarks about this event can be found in \S5.

\section{Swift J1644+57: Summary of Observations and puzzles they have posed}

Swift/XRT found that the x-ray flux from Sw J1644+57 varied rapidly on 
time scale $\delta t_{obs}\sim10^2$s, indicating that the emission was 
probably produced at a distance of order $2c\delta t_{obs} \Gamma_j^2\lae
10^{15}$cm; $\Gamma_j \lae 10$ is jet Lorentz factor. The average x-ray 
flux was roughly constant for the first 5 days, then dropped suddenly 
by a factor 50 during the next 3 days, and subsequently for $20{\rm d}\lae 
t_{obs}\lae500$d it declined as $\sim t_{obs}^{-5/3}$ (Burrows et al. 2011; 
Berger et al. 2012; Zauderer et al. 2013).
The x-ray flux then declined further by a factor $\sim 170$ between 500 days
and 600 days (Zauderer et al. 2013). The spectrum in the 0.3--10 keV
Swift/XRT band is reported to be a power law function --- $f_\nu \propto
\nu^{-\beta}$ --- with index $\beta\sim 0.7$ during the first
$\sim50$days and then decreasing to $\sim 0.4$ a few hundred days
after the tidal disruption. 

This transient event received extensive coverage at multiple mm and radio
frequencies. Depending on the frequency band the light curve was found to rise,
or remained nearly flat, during the first 100 days, and then declined as 
roughly $t_{obs}^{-1}$. The spectrum at low frequencies ($\lae 10^{10}$Hz) 
rose as $\sim\nu^{1.5}$ which is roughly consistent with that expected of 
a synchrotron spectrum below the self-absorption frequency ($\nu_a$). And 
the spectrum above the peak of $f_\nu$ fell off as $\sim\nu^{-0.7}$. The 
synchrotron self-absorption frequency ($\nu_a$) underwent a dramatic decline 
by a factor of 10 between 5 and 10 days, and subsequently $\nu_a$ displayed a 
very slow decline over the next several hundred days. The flux at the peak
of the spectrum fell off during 5--10 days by a factor $\sim 3$, then
increased by a factor $\sim 4$ during 10--10$^2$ days (Berger et al. 2012).
It should be noted that these peculiar behavior of the radio data during the
period 5--10 days coincided with a sharp decline of the x-ray flux by a
factor $\sim10^2$, e.g.  Burrows et al. (2011), Berger et al. (2012).

The event was also observed in the infrared K-band ($\sim 1.4\times10^{14}$Hz)
multiple times between 5 and 50 days, and the flux was found to decline during 
the first 5 days as $\sim t_{obs}^{-3}$ (Levan et al. 2011).

One of the main puzzles posed by the rich data for this event is that 
light curves in $\gae10$GHz frequency bands, which lie above the 
synchrotron absorption and other characteristic frequencies, show a flat 
or a rising behavior when in fact according to the external forward shock 
model these light curves should be declining as $\sim t_{obs}^{-1}$, e.g. 
Meszaros \& Rees (1994), Chevalier and Li (2000), Panaitescu \& Kumar (2000), 
Granot and Sari (2002). Energy injection to the decelerating external 
blast wave between 10 and $10^2$ days by a factor $\sim 20$ has been 
invoked, e.g. Berger et al. (2012), to prevent the decline of the radio 
light curve during this time interval. 
This huge amount of energy 
added to the blast wave, however, leaves no x-ray foot print, and
that clearly is very puzzling\footnote{The x-ray light curve fell off as 
$t_{obs}^{-5/3}$ for $t_{obs}\gae 5$d, and therefore according to the
x-ray data most of the energy release in relativistic jet took place 
during the first 5 days.}. The model we suggest --- IC cooling
of electrons by x-ray photons --- avoids this problem because it
requires no energy addition to the decelerating external shock 
to flatten radio/mm light curves.

Two other puzzles suggested by the data are the precipitous decrease of 
$\nu_a$ by a factor 10 between 5 and 10 days, and the rapid fall off of 
the infrared light curve during the same period.

\section{Influence of x-rays on External shock radiation}

\noindent{\bf Basic idea:} relativistic jet from the TDE interacts
with the CNM (circum-nuclear medium) and drives a strong shock wave
into it\footnote{A reverse shock propagating into the jet can also be produced 
in the CNM--jet interaction provided that the jet is not Poynting flux dominated, 
i.e. the ratio of energy flux carried by magnetic fields and the kinetic energy
of matter ($\sigma$) is not much larger than 1. The mechanism described in this
paper can also be applied to reverse shock emission. However, the external
reserve-shock has difficulties explaining the observed flat radio/mm light curves.}. 
CNM electrons accelerated by the shock radiate synchrotron photons at radio and
higher frequencies. These electrons are cooled by the IC scattering of x-ray 
photons streaming through the shocked plasma\footnote{X-rays observed by the Swift
satellite must have passed through the external shock region as long as 
they are produced at the same, or smaller, radius as the external
shock. The very high variability of the x-ray data -- on timescale of
 order minutes --- suggests that the x-ray radiation is likely produced at 
a distance much smaller than the external shock radius.}
 (these x-ray photons are the same radiation that was observed by the
Swift satellite); see Fig. 1 for a schematic sketch of this scenario.
Since the x-ray flux declines rapidly with time ($f_x \propto t^{-5/3}$
for $t\gae 10$ days), the IC cooling of electrons also diminishes with time
and this is what is responsible for transforming an otherwise declining
light curve ($f_\nu\aprop t_{obs}^{-1}$) to a flat or a rising light curve in
radio/mm band;
the IC cooling by x-rays turns out to be very effective in suppressing 
synchrotron radiation in mm/cm bands at early times, and this suppression
gets weaker with time as the x-ray flux declines. 

 One other thing we should address here for those who might be thinking that 
the observed spectrum above the peak ($f_\nu\aprop \nu^{-0.8}$) is too 
shallow to be consistent with a fast electron cooling model. The observed 
parameters of Swift J1644+57 are such that IC scatterings for electrons 
radiating above the peak takes place in the Klein-Nishina regime, and hence the
shallow observed spectrum; detailed calculations are presented in \S3.1 \& 3.2.

\begin{figure}
\centerline{\hbox{\includegraphics[width=9cm, angle=0]{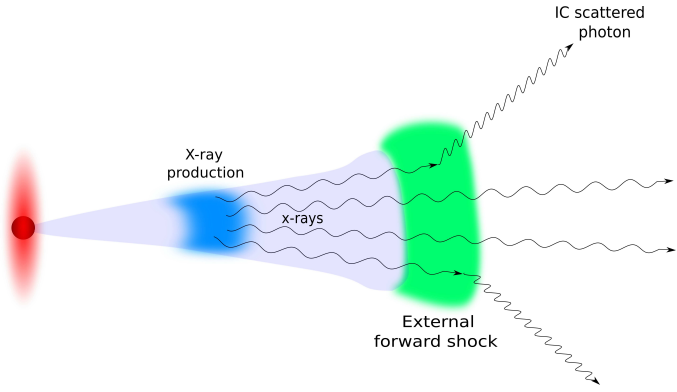}}}
\caption{Shown here is a schematic sketch of a jet produced in TDE. The 
energy is converted into x-rays at some distance from the black-hole by an as 
yet unknown mechanism. A part of this x-ray radiation is scattered by 
electrons in the forward-external shock which results in their efficient 
cooling at early times.  
}
\label{fig1}
\end{figure}

Since the main component of our proposed idea to explain radio/mm light curve
behavior is IC cooling of electrons in the external shock by the 
observed x-ray radiation, we begin the technical part of this section 
with a description of this process. We show that it is unavoidable that 
IC cooling by x-ray photons significantly modifies electron distribution 
in the external shock for Sw J1644+57. We also discuss whether IC scatterings 
take place in the Thomson or Klein-Nishina (K-N for short, from now on) regime.
Finally, we calculate electron distribution function when IC cooling due to an
external x-ray radiation field occurs in the K-N regime.

\subsection{Importance of IC cooling and its effect on electron distribution}

Let us consider a x-ray radiation source with luminosity $L_x(t_{obs})$ 
(isotropic equivalent) in frequency band $\nu_1$--$\nu_2$ which is produced 
by a source of size $r_x$. Its specific luminosity scales as $L_\nu
\propto \nu^{-\beta}$; all unprimed variables are quantities measured in 
the rest frame of the host galaxy of the source. Now consider electrons in 
external forward shock
at a radius $R(t_{obs})$, and the Lorentz factor of the shocked plasma
(wrt CNM rest frame) to be $\Gamma(t_{obs})$. A photon of 
frequency $\nu$ moving at an angle $\theta$ wrt the radial direction 
is Doppler shifted to frequency $\nu'$ in shock comoving frame:
\begin{equation}
  \nu' = \nu\Gamma(1 - \beta \cos\theta)\approx {\nu\over 2\Gamma}
       \left[1 + \theta^2\Gamma^2\right], 
  \label{nup}
\end{equation} 
where
\begin{equation}
    \beta = \left(1 - \Gamma^{-2}\right)^{1/2},
\end{equation}
and we used Taylor expansion for $\cos\theta$, and $\Gamma\gg1$ to
arrive at the approximate expression for $\nu'$. From hereon
all primed variables denote shock comoving frame. 

The specific intensity in the host galaxy rest frame is
\begin{equation}
   I_\nu(\theta) \approx {L_\nu\over 4\pi^2 r_x^2},
\end{equation}
for $\theta \lae r_x/R\equiv \theta_x$, and is zero otherwise.
The specific intensity in the comoving frame of the shocked fluid is
obtained by an appropriate Lorentz transformation, e.g. Rybicki 
\& Lightman (1986), and is given by
\begin{equation}
   I'_{\nu'} = I_\nu (\nu'/\nu)^3 \sim I_\nu/(8\Gamma^3),
\end{equation}
when $\theta\ll \Gamma^{-1}$.  Therefore, the specific flux in the comoving 
frame is
\begin{equation}
   f'_{\nu'} \sim (\pi \theta_x'^2) I'_{\nu'} \sim {L_\nu\over 8\pi\Gamma R^2},
\end{equation}
where $\theta'_x \sim 2\Gamma\theta_x\ll 1$ is the angular size of the x-ray
source as seen in the shock comoving frame. And the frequency integrated flux is
\begin{equation}
   f'_x \sim {L_x\over 16\pi \Gamma^2 R^2}.
  \label{fxp}
\end{equation}
The factor $\Gamma^2$ in the denominator is easy to understand as it 
is due to Lorentz transformation of energy density.

The equation for IC cooling of an electron of Lorentz factor $\gamma_e'$ is
\begin{equation}
   {d (m_e c^2 \gamma_e')\over dt_{ic}'} = -{4\over 3} \sigma_T f_x' \gamma_e'^2,
  \label{gam_dot}
\end{equation}
where $\sigma_T$ is Thomson scattering cross section. At the moment we are
ignoring K-N effect which we will take up shortly.
Thus, IC cooling time is given by
\begin{equation}
   t_{ic}' \sim {m_e c^2\over \sigma_T f_x' \gamma_e'} \sim (6\times10^6 s)
   R_{17}^2 L_{x,47}^{-1} (\Gamma^2/\gamma_e'),
\end{equation}
where we made use of equation (\ref{fxp}) for $f_x'$, and have adopted 
the widely used notation $X_n \equiv X/10^n$. The IC cooling calculation 
assumed that the external shock region is optically thin for x-ray photons.
This condition is easily satisfied since the optical depth to 
Thomson scattering is $\sim \sigma_T n_e R \sim 10^{-7} n_e R_{17} \ll 1$
as long as $n_e$, the particle density in the unshocked CNM, is less than
10$^7$cm$^{-3}$ at a distance of 10$^{17}$cm from the black hole which
is indeed the case as shown by radio observations. Also, absorption of
x-ray photons by the inverse-Synchrotron process is negligibly small.

The dynamical time in the comoving frame is $t'_{dy} \sim R/(c\Gamma)$.
Therefore,
\begin{equation}
   {t_{ic}'\over t_{dy}'} \sim {2 \Gamma^3 R_{17} \over \gamma_e' L_{x,47} }.
   \label{tic_tdy}
\end{equation}
The Lorentz factor of the external shock a week after the tidal
disruption is about 2 (see \S3.2 for shock dynamics),
$R_{17} \sim 1$, and $L_x\sim 5$x$10^{46}$erg/s. Thus we see from the above
equation that the IC cooling time for electrons with $\gamma_e' \gae 30$
is shorter than the dynamical time. Radio and mm photons observed from
Sw J1644+57 are produced via synchrotron process by electrons with 
$\gamma_e' \gae 50$, and so IC cooling cannot be ignored in the 
calculation of external shock radiation from this event. 
It should be pointed out that the synchrotron cooling is less important
than the IC cooling at least for those electrons that are radiating 
well below the infrared band\footnote{The ratio of IC and synchrotron cooling
times, in Thomson scattering regime, is equal to the ratio of energy density 
in magnetic field and radiation field. The x-ray luminosity from Sw J1644+57 
was very high for the first several weeks, and thus it turns out that the 
energy density in radiation is much larger than magnetic energy density as 
long as $\epsilon_B$ for the shocked fluid is smaller than 1.}.

For electrons with $\gamma_e' \lae 10^2$ the IC scatterings take place
in the Thomson regime for Sw J1644+57. The reason is that the energy of a
10 keV photon --- Swift/XRT energy band is 0.3--10 keV --- 
in the electron comoving frame is ($ 10 \gamma_e'/\Gamma$) keV
which is less than $m_e c^2$ as long as $\gamma_e'\lae 10^2$. 
Higher energy electrons scatter 10 keV photons with a reduced
cross-section\footnote{Klein-Nishina cross-section is smaller than 
the Thomson cross section by a factor of photon energy in electron rest 
frame divided by $m_e c^2$.}, and the scattered photon
energy is less than it is in the Thomson regime, i.e. $<\nu'\gamma_e'^2$.
These two effects reduce the IC cooling of electrons with $\gamma_e'\gae10^2$.
A simple, and fairly accurate, way to include these effects in IC cooling 
calculations is by only considering energy density in radiation field 
up to a photon energy of $\epsilon_{kn}'$:
\begin{equation}
  \epsilon_{kn}'\sim m_e c^2/\gamma_e'  \quad{\rm or}\quad \nu_{kn}' = \epsilon_{kn}'/h,
   \label{nu_kn}
\end{equation}
where $h$ is Planck's constant.
The spectral index in the x-ray band for Swift J1644+57 is $\beta\approx 0.7$,
and in that case the reduction to the IC cooling due to K-N effect 
scales as $\gamma_e'^{-0.3}$ for large $\gamma_e'$ (see \S3.1.1). This is a 
weak effect but has important implications for electron distribution function
which is discussed below.

\subsubsection{Electron distribution function in external shock}

Electron distribution function, $d n_e/d\gamma_e$, in a shocked plasma 
is determined by the dual effect of acceleration at the shock front and 
IC and synchrotron coolings while electrons travel down-stream. The 
distribution function is obtained from the following equation:
\begin{equation}
    {\partial \over \partial t}{d n_e\over d\gamma_e} + {\partial
    \over \partial \gamma_e} \left[ \dot{\gamma_e} {d n_e\over
   d\gamma_e}\right] = S(\gamma_e),
  \label{ne_dot}
\end{equation}
where 
\begin{equation}
  \dot{\gamma_e} = - {4\sigma_T\over 3 m_e c^2} \left[ f(<\nu_{kn}) + 
   {B^2 c\over 8\pi}\right] \gamma_e^2\beta_e^2,
   \label{gamma_e_dot}
\end{equation}
is the rate of change of $\gamma_e$ due to IC and synchrotron losses, 
$f(<\nu_{kn})/c$ is energy density in radiation below the K-N frequency 
$\nu_{kn}$ (eq. \ref{nu_kn}), and 
\begin{equation}
   S(\gamma_e)\propto \gamma_e^{-p} \quad {\rm for}\;\, \gamma_e\ge\gamma_i
\end{equation}
 is the rate at which electrons with LF $\gamma_e$ are injected into the 
system; $\gamma_i$ is the minimum Lorentz factor for shock accelerated 
electrons. All the variables considered in this subsection are in shock
comoving frame. However, we will not use prime (') on variables in this
sub-section (and only in this sub-section) to denote shock comoving frame 
as this is the only frame being considered here. When results from this
sub-section are used elsewhere in the paper we will revert back to
the proper notation for prime and un-prime frames.

Let us define radiative cooling time
\begin{equation}
    t_{c}(\gamma_e) = {\gamma_e\over \dot{\gamma_e} },
\end{equation}
and cooling Lorentz factor, $\gamma_c$, which is such that 
\begin{equation}
    t_{c}(\gamma_c) \equiv t_{dy},
\end{equation}
where $t_{dy}$ is the dynamical time.

We will calculate the distribution function in two cases. The slow cooling
case where $\gamma_c > \gamma_i$, and the fast cooling regime when 
$\gamma_c < \gamma_i$. Moreover, we will consider here only IC losses in
K-N regime as that is relevant for the ``afterglow'' radiation of the TDE 
Swift J1644+57. We note that the theoretical light curves and spectra 
shown for this event in the next section are obtained by numerical 
solutions of relevant equations that include synchrotron loss and no 
assumption regarding K-N regime.

Since $\nu_{kn}\propto \gamma_e^{-1}$ (eq. \ref{nu_kn}) and $f(<\nu_{kn})
\propto \gamma_e^{-(1-\beta)}$, therefore $\dot{\gamma_e}\propto
\gamma_e^{1+\beta}$. We see from equation (\ref{ne_dot}) that 
for $\gamma_e < \gamma_i$
\begin{equation}
   {d n_e\over d\gamma_e} \propto {1\over \dot{\gamma_e}}
\end{equation}
whereas for $\gamma_e > \gamma_i$
\begin{equation}
   {d n_e\over d\gamma_e} \propto {\int_{\gamma_e}^\infty d\gamma 
   S(\gamma)\over \dot{\gamma_e} }.
\end{equation}
Therefore, the solution of equation (\ref{ne_dot}) in the slow cooling
case is
\begin{equation}
{d n_e\over d\gamma_e} \propto \left\{
\begin{array}{ll}
  \gamma_e^{-p}   \quad\quad  & \gamma_i\le \gamma_e \le \gamma_c \\
\gamma_e^{-p-\beta} \quad\quad \quad\quad & \gamma_e > \gamma_c
\end{array}
\right.
  \label{dne_dgam3}
\end{equation}
and in the fast cooling case
\begin{equation}
{d n_e\over d\gamma_e} \propto \left\{
\begin{array}{ll}
  \gamma_e^{-1-\beta}   \quad\quad  & \gamma_c\le \gamma_e \le \gamma_i \\
\gamma_e^{-p-\beta} \quad\quad\quad\quad & \gamma_e > \gamma_i
\end{array}
\right.
  \label{dne_dgam4}
\end{equation}

The synchrotron spectrum for particle distribution $d n_e/d \gamma_e 
\propto \gamma_e^{-q}$ is $f_\nu\propto \nu^{-(q-1)/2}$ for $\nu > \min(\nu_c,
\nu_i)$, and it is $\nu^{1/3}$ for $\nu < \min(\nu_c, \nu_i)$; these spectra
are modified when synchrotron-absorption becomes important (the criterion 
for this is discussed in the next sub-section)\footnote{A very detailed 
description of synchrotron spectrum when IC cooling in K-N regime is
important can be found in Nakar et al. (2009)}. 

It should be clear from these equations that the spectral index above
the peak depends on $p$, and the spectral index $\beta$ of the radiation field
that is responsible for IC cooling of electrons; the spectral index for 
$\nu > \max(\nu_i, \nu_c)$ is $(p-1+\beta)/2$ which is harder than 
$\nu^{-p/2}$. As an application to Swift J1644+57 we note that the spectral 
index in the radio band above the peak of the spectrum is reported to be 
$\sim 0.75$ for $t_{obs}\gae 10^2$d (Berger et al. 2012). This 
does not mean that $p=2.5$ as suggested by eg. Berger et al.
If IC cooling in K-N regime is important for determining particle distribution,
which equation (\ref{tic_tdy}) for $t'_{ic}/t_{dy}'$ makes clear is the case, 
then $p=2.3$, since according to the late time X-ray data $\beta = 0.4$ 
(Saxton et al. 2012).

\subsection{Shock dynamics and synchrotron light curves}

The dynamics of external forward shock can be determined, approximately,
from the following energy conservation equation
\begin{equation}
   {4\pi \over (3-s)} R^3 R_{17}^{-s} n_{f} m_p c^2 \Gamma(\Gamma-1) = E,
  \label{E_conserve}
\end{equation}
where $E$ is the energy (isotropic equivalent) in the blast wave, the density 
of the CNM is taken to be 
\begin{equation}
   n(R) = n_{f} R_{17}^{-s},
   \label{nR}
\end{equation}
and $R_{17} = R/10^{17}$cm. For a relativistic shock, i.e. $\Gamma\gg1$,
the dynamics is described by
\begin{equation}
\Gamma = \left[ { (3-s) E_{51}\over 4\pi m_p c^2 n_{f} R_{17}^{3-s} }
      \right]^{1/2},
\end{equation}
and the deceleration radius -- where the relativistic ejecta has imparted
roughly half its energy to the surrounding medium -- is
\begin{equation}
   R_{d,17} = \left[ {(3-s) E_{51} \over 4\pi m_p c^2 n_{f} \Gamma_0^2} 
   \right]^{1/(3-s)},
\end{equation}
where $\Gamma_0$ is the initial Lorentz factor of the relativistic outflow.
For a non-relativistic shock the solution of equation (\ref{E_conserve}) is
\begin{equation}
  v = \left[ { (3-s) E_{51}\over 2\pi m_p n_{f} R_{17}^{3-s} }
      \right]^{1/2}.
\end{equation}
The Lorentz factor for the external shock of TDE Swift J1644+57 was of order
a few at early times (less than a week), and the shock speed decreased to 
Newtonian regime at late times. Hence the analytical calculations we present
in this section considers both of these regimes.  The numerical results
described in the next section are more accurate and do not rely on making
any assumption regarding $\Gamma$.

A photon released at radius $R$ arrives at the observer at time
\begin{equation}
   t_{obs} = {(1+z)\over c} \int dR \left[ {c\over v} - 1\right].
   \label{tobs}
\end{equation}
For relativistic blast waves $t_{obs}\propto R^{4-s}$, and for 
non-relativistic case $t_{obs}\propto R^{(5-s)/2}$. Moreover, the
shock speed  as viewed by an observer is
\begin{equation}
   \Gamma \propto t_{obs}^{-{3-s\over 8-2s}} \quad\quad {\rm and} \quad\quad
            R \propto t_{obs}^{1/(4-s)},
   \label{dyna1}
\end{equation}
for the relativistic case and
\begin{equation}
   v \propto t_{obs}^{-{3-s\over 5-s}} \quad\quad {\rm and} \quad\quad
            R \propto t_{obs}^{2/(5-s)},
   \label{dyna2}
\end{equation}
for Newtonian dynamics.

The deceleration time in observer frame, for relativistic outflow, is given by
\begin{equation}
   t_d \approx (3\times10^4 {\rm s}) \left[ { (3-s) E_{53}\over 1.8\,n_{f,2}}
       \right]^{ {1\over 3-s}} (1+z) \Gamma_{0,1}^{-{8-2s\over 3-s}}.
\end{equation}
Thus, for $E=10^{53}$erg, $n_f=10^2$cm$^{-3}$, and $\Gamma_0=10$, the 
deceleration time is of order 5 hours, and the Lorentz factor of the 
shocked CNM 10 days after the start of TDE is $\sim2$.

\subsubsection{External shock radiation and light curves}

We provide analytical calculation for synchrotron radiation and light curves 
from external forward shock in this sub-section; more accurate numerical 
results are presented in the next section. 

The magnetic field strength and the electron minimum Lorentz factor in
a shock heated plasma are estimated by assuming that a fraction of the energy 
of shocked plasma ($\epsilon_B$) goes into the generation of magnetic fields, 
and a fraction ($\epsilon_e$) is taken up by electrons\footnote{This
crude approximations is made due to 
our inability to calculate magnetic fields and electron acceleration
in collisionless shocks ab initio. And the dimensionless parameters 
$\epsilon_B$ and $\epsilon_e$ hide our ignorance regarding these 
processes.}. Under this assumption we find that
\begin{equation}
    B' = \left[ 32\pi \epsilon_B m_p c^2 n(R) \Gamma (\Gamma-1)\right]^{1/2},
   \label{Bp}
\end{equation}
and the minimum Lorentz factor of shock accelerated electrons is
\begin{equation}
  \gamma_i' = {(p-2)m_p\over (p-1)m_e} \epsilon_e (\Gamma -1) 
   \label{gamma_i}
\end{equation}

Using results for shock dynamics from last sub-section we find
\begin{equation}
   B' \propto \left\{
\begin{array}{ll}
  \hskip -5pt  t_{obs}^{-{3\over 2(4-s)}} \quad\quad  &  \Gamma\beta\gg 1\\ \\
 \hskip -5pt t_{obs}^{-{3\over 5-s}}  \quad\quad\quad\quad\quad\quad\quad\quad &  \Gamma\beta \ll 1
\end{array}
\right.
  \label{Bp1}
\end{equation}
and the synchrotron frequency corresponding to electron Lorentz factor 
$\gamma_i'$ is
\begin{equation}
    \nu_i = {q B' \gamma_i'^2 \Gamma \over 2\pi m_e c (1+z)} \propto \left\{
\begin{array}{ll}
  \hskip -5pt t_{obs}^{-3/2}  \quad\quad  & \Gamma\beta\gg 1 \\ \\
 \hskip -5pt t_{obs}^{-{15-4s\over 5-s} }\quad\quad & \Gamma\beta \ll 1
\end{array}
\right.
   \label{nui}
\end{equation}

We next calculate synchrotron frequency for those electrons that cool
on a dynamical timescale ($\nu_c$) due to IC scattering of x-ray photons.
Let us rewrite equation (\ref{gam_dot}) to explicitly consider x-ray flux
below the K-N frequency
\begin{equation}
   {d m_e c^2 \gamma_e'\over dt_{ic}'} = -{4\over 3} \sigma_T f_x'(<\nu_{kn}')
    \gamma_e'^2, \label{gam_dot1}
\end{equation}
where $\nu_{kn}'$ is given by equation (\ref{nu_kn}) for electrons of 
LF $\gamma_c'$. The LF of electrons that 
cool on a dynamical time is obtained from the above equation
\begin{equation}
   \gamma_c' \sim {3 m_e c^2 \over 4 \sigma_T f_x'(<\nu_{kn}') t_{dy}'}.
\end{equation}
For a power law spectrum for x-ray radiation
\begin{equation}
   f_x'(<\nu_{kn}') = {L_x(<\nu_{kn})\over 16\pi \Gamma^2 R^2} \propto
      {L_x(t_{obs}) \over R^2 \Gamma^{1+\beta} \gamma_c'^{1-\beta} }.
\end{equation}
Therefore, the electron ``cooling'' Lorentz factor is given by
\begin{equation}
    \gamma_c' \propto \left[ { R^2 \Gamma^{1+\beta} \over t_{dy}' L_x(t_{obs})}
      \right]^{1/\beta},
\end{equation}
Using equations for shock dynamics from last sub-section we find
\begin{equation}
  \gamma_c' \propto \left\{
\begin{array}{ll}
    \left[ {t_{obs}^{[2-(2+\beta)(3-s)]/[2(4-s)]}\over L_x(t_{obs}) 
        }\right]^{1/\beta} \quad & {\rm for}\;\, \Gamma\beta\gg 1 \\ \\
    \left[ {t_{obs}^{(s-1)/(5-s)}\over L_x(t_{obs})} \right]^{1/\beta}
         \quad & {\rm for}\;\, \Gamma\beta\ll 1
\end{array}
\right.
\end{equation}
The synchrotron frequency in
observer frame corresponding to $\gamma_c'$ is
\begin{equation}
  \nu_c = {q B' \gamma_c'^2 \Gamma\over 2\pi (1+z) m_e c^2}.
   \label{nuc2}
\end{equation}
Or
\begin{equation}
  \nu_c \propto \left\{
\begin{array}{ll}
   \hskip -5pt t_{obs}^{-3/2} \left[ {t_{obs}^{(s-2)/(4-s)}\over L_x(t_{obs}) }
        \right]^{2/\beta} \quad & {\rm for}\;\, \Gamma\beta\gg 1 \\  \\
   \hskip -5pt t_{obs}^{-3/(5-s)} \left[ {t_{obs}^{(s-1)/(5-s)}\over 
    L_x(t_{obs})} \right]^{2/\beta} \quad & {\rm for}\;\, \Gamma\beta\ll 1
\end{array}
\right.
   \label{nuc3}
\end{equation}
For $s=1.5$, $\beta=0.7$ and $L_x\propto t_{obs}^{-5/3}$ we find $\nu_c
\propto t_{obs}^{2.7}$ for relativistic case, and $\nu_c\propto
t_{obs}^{4.3}$ for Newtonian dynamics. For these parameters for Sw J1644+57,
$\nu_c$ is about 1 GHz at 10 days, and it rises to 100 GHz at $\sim70$ days.

The observed flux at a frequency $\nu > \max(\nu_a, \nu_i, \nu_c)$ is given
by
\begin{equation}
 f_\nu(t_{obs}) = f_p \nu_c^{\beta/2} \nu_i^{(p-1)/2}\nu^{-(p+\beta-1)/2},
   \label{fnu}
\end{equation}
where $\nu_a$ is synchrotron absorption frequency, 
\begin{equation}
   f_p  \approx {q^3 B' \Gamma N_e \over m_e c^2} {1+z\over 4\pi d_L^2},
   \label{fp}
\end{equation}
$N_e\propto R^{3-s}$ is the total number of electrons with Lorentz factor 
$>\min(\gamma_i', \gamma_c')$, and $d_L$ is the luminosity distance to the 
source; we made use of equations (\ref{dne_dgam3}) and (\ref{dne_dgam4}) in 
deriving equation (\ref{fnu}).

Making use of equations (\ref{dyna1}), (\ref{dyna2}), (\ref{Bp1}),
(\ref{nui}) and (\ref{nuc3}) in the above expression for $f_\nu$ we find 
the observed light curve behavior:
\begin{eqnarray}
  && f_\nu(t_{obs}) \propto{\nu^{-(p+\beta-1)/2}\over L_x(t_{obs})}\;\;\times 
   \nonumber \\
   && \quad\quad\quad\quad\quad\quad \left\{
\begin{array}{ll}
  \hskip -5pt t_{obs}^{-[2+3\beta+3(p-1)]/4}  \quad & \Gamma\beta\gg 1 \\ \\
  \hskip -5pt t_{obs}^{ {19-6s-p(15-4s)-3\beta\over [2(5-s)]}} \quad\quad\;
     & \Gamma\beta\ll 1
\end{array}
\right.
  \label{fnu2}
\end{eqnarray}
For $p=2.0$, $\beta=0.7$, $s=1.5$ and $L_x \propto t_{obs}^{-5/3}$, 
$f_\nu \propto \nu^{-0.85} t_{obs}^{-0.1}$ for relativistic dynamics,
and $f_\nu\propto \nu^{-0.85} t_{obs}^{0.2}$ for Newtonian shock.
The dynamics of Swift J1644+57 for $10{\rm d} \lae t_{obs} \lae 10^2$d 
lies between these two extremes, and the light curve is flat for as long
a duration as the observing stays above $\nu_c$. Of course, at sufficiently
high frequencies --- which of order $10^{13}$Hz for this TDE at
$\sim 10$d --- electron cooling by IC scattering becomes unimportant
due to K-N suppression, and synchrotron cooling time becomes smaller
than the dynamical time. In this case the light curve decline has the
well known form of $f_\nu\propto \nu^{-p/2} t_{obs}^{-(3p-2)/4}$
for relativistic shocks, e.g. Kumar (2000).

The rapid increase of $\nu_c$ with time is responsible for the nearly
flat light curve observed for Swift J1644+57 in frequency bands 
$\nu > \max(\nu_a, \nu_i, \nu_c)$. When $\nu_c(t_{obs})$ moves above the
observing band then the light curve resumes its usual, $\sim t_{obs}^{-1}$,
decline which occurs at $t_{obs}\gae10^2$days for $\nu\gae10^{2}$GHz as 
shown in numerically calculated light curves presented in the next section.

Another interesting case to consider is when $\nu_c < \nu < \nu_i$
and the observing band is above the synchrotron absorption
frequency. The light curve in this case is determined from
\begin{equation}
   f_\nu(t_{obs}) = f_p \nu^{-\beta/2} \nu_c^{\beta/2}.
\end{equation}
Making use of equations (\ref{fp}) and (\ref{nuc3}) we find
\begin{equation}
f_\nu(t_{obs}) \propto {\nu^{-\beta/2}\over L_x(t_{obs}) } \left\{
\begin{array}{ll}
  \hskip -5pt t_{obs}^{-(2+3\beta)/4}   & \Gamma\beta\gg 1 \\ \\
  \hskip -5pt t_{obs}^{[4-2s-3\beta]/[2(5-s)]} \quad\quad & \Gamma\beta\ll 1
\end{array}
\right.
\end{equation}
For $\beta=0.7$, $s=1.5$ and $L_x\propto t_{obs}^{-5/3}$ the light curve 
rises as $\nu^{0.35}\,t^{0.64}$ ($t_{obs}^{1.5}$) for relativistic (Newtonian)
blast waves. 

Next, we calculate the synchrotron self-absorption frequency ($\nu_a$)
when IC cooling is important. The easiest way to determine $\nu_a$ is by 
using Kirchhoff's law or the fact that the emergent flux at
frequency $\nu_a$ is approximately equal to the black-body flux
corresponding to electron temperature given by
\begin{equation}
   k T_e' = {\rm max} [\gamma_a', {\rm min}(\gamma_i',\gamma_c')] m_e c^2/3,
   \label{Ta}
\end{equation}
where $\gamma_a'$ is electron Lorentz factor corresponding to synchrotron 
frequency $\nu_a'$, i.e. $\gamma_a'\propto (\nu_a'/B')^{1/2}$. The equation for 
$\nu_a'$, using Kirchhoff's law, is
\begin{equation}
   {2 k T_e' \nu_a'^2 \over c^2} \approx
         {\sigma_T B' m_e c^2 N_e(>k T_e/m_e c^2)\over 4\pi q},
  \label{nua1}
\end{equation}
where the left side of this equation is the Planck function in the
Rayleigh-Jeans limit, and $N_e(>k T_e/m_e c^2)$ is the column density
of electrons with LF larger than ${\rm max}[\gamma_a',{\rm min}
(\gamma_i',\gamma_c')]$.

One case of particular interest for application to Sw J1644+57 is where
$\nu_a$ is larger than $\max(\nu_i, \nu_c)$. We can rewrite equation
(\ref{nua1}) for this case as
\begin{equation}
   2 m_e \nu_a'^{5/2} \left( {2\pi m_e c \over q B'}\right)^{1/2} =
     f'(\nu_a') = {f_{\nu_a} d_L^2 \over \Gamma R^2}. 
   \label{nua2}
\end{equation}
Making use of equations (\ref{dyna1}), (\ref{dyna2}), (\ref{Bp1}) and
(\ref{fnu2}) we find
\begin{equation}
\nu_a \propto \left\{
\begin{array}{ll}
  \hskip -5pt {t_{obs}^{-{8+(4-s)(3p+3\beta+2)\over 2(4-s)(p+\beta+4) } }\over
       L_x^{{2\over p+\beta+4}} }  & \Gamma\beta\gg 1 \\ \\
  \hskip -5pt {t_{obs}^{-{30 - 6s - p(15-4s) - 3\beta\over (5-s)(p+\beta+4) }}
    \over  L_x^{{2\over p+\beta+4}} }  \quad\quad\quad & \Gamma\beta\ll 1
\end{array}
\right.
   \label{nua3}
\end{equation}
For $s=1.5$, $p=2$, $\beta=0.7$, and $L_x\propto t_{obs}^{-5/3}$, 
$\nu_a\propto t_{obs}^{-0.49}$ ($t_{obs}^{0.45}$) for relativistic
(Newtonian) shock dynamics. We note that due to the rapid rise of 
$\nu_c$ (see eq. \ref{nuc3}) the above equation for $\nu_a$ ceases to be
applicable after a few tens of days for Sw J1644+57 when the ordering of 
synchrotron characteristic frequency changes to $\nu_i < \nu_a < \nu_c$.
The synchrotron absorption frequency for this case can be shown to be
\begin{equation}
    \nu_a \propto \left\{
\begin{array}{ll}
  \hskip -5pt t_{obs}^{- {1\over p+4}\left[{10-s\over 2(4-s)} + {3(p-1)\over 2}
     \right]}   & \Gamma\beta\gg 1 \\ \\
  \hskip -5pt t_{obs}^{-{\left[(5+4s)+(15-4s)(p-1)\right]\over (5-s)(p+4)}
    }  \quad\quad\quad  &    \Gamma\beta\ll 1
\end{array}
\right.
\end{equation}
For $s=1.5$ and $p=2$, $\nu_a\propto t_{obs}^{-0.53}$ ($t_{obs}^{-0.95}$) 
for relativistic (Newtonian) shock dynamics.

The rise of light curve for about $10^2$days for $\nu\lae10$GHz is due 
to the fact that these observing bands lie below the synchrotron absorption
 frequency during this period. The observed flux below $\nu_a$ decreases rapidly 
with decreasing $\nu$, and hence even a slow decrease of $\nu_a$ with
time leads to a rising light curve.

\section{Numerical results and application to Swift J1644+57}

Equations (\ref{E_conserve}) \& (\ref{tobs}) are solved numerically to 
determine shock dynamics as a function of observer time, and
equations (\ref{ne_dot}), (\ref{gamma_e_dot}), (\ref{Bp}), (\ref{gamma_i})
\& (\ref{nua1}) are used for calculating spectra and light curves that 
take into account electron cooling due to IC scattering
of an ``external'' radiation field with flux and spectrum taken from x-ray 
observations, and also cooling due to synchrotron radiation.
No assumption regarding whether IC scatterings are in K-N or Thomson regime
are made for calculations presented in this section.

The energy in the blast wave is injected only at the beginning of numerical
calculation when the blast wave is at a radius $R=10^{13}$cm; results are 
insensitive to the precise value of the initial radius as long as it is 
taken to be much smaller than the deceleration radius.
In Fig. 2 we show $R$, $\Gamma$, $B'$, $\gamma_i$ \& $\gamma_c$,
for parameters relevant for Sw J1644+57. 

\begin{figure}
\centerline{\hbox{\includegraphics[width=9cm, angle=0]{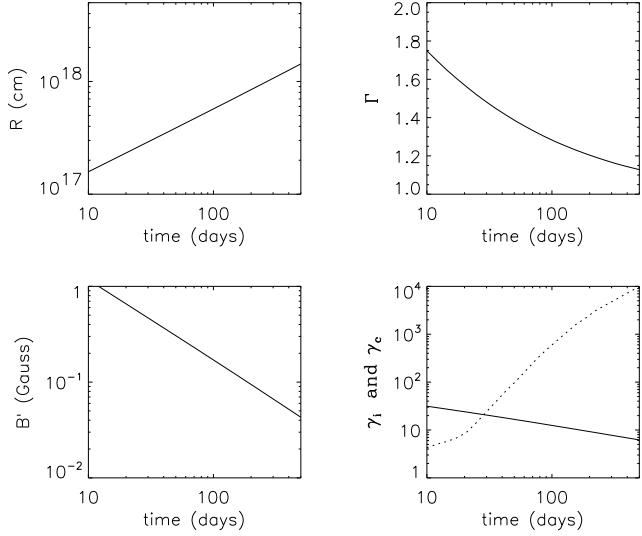}}}
\caption{A few of the parameters of the external forward show are shown in this
figure as a function of observer time. Shock front radius (top left), Lorentz
factor of the shocked fluid (top right), magnetic field in the comoving frame 
of shocked fluid (bottom left), and bottom right panel shows $\gamma_i$ 
(solid line) \& $\gamma_c$ (dotted line). The energy in the blast wave was 
taken to be $E=1.5\times10^{53}$erg (isotropic), the CNM density 
$n(R)=(400{\rm cm}^{-3}\times R_{17}^{-2}$, $\epsilon_B = 
4.5\times10^{-3}$, $\epsilon_e = 0.1$, and $p=2.3$ for these calculations. 
The x-ray luminosity was taken to be $L_x = 4\times10^{46}{\rm erg/s} 
(t_{obs}/20 {\rm days})^{-5/3}$ for $t_{obs}>20$d and constant between 
10 and 20 days; x-ray photons passing through
the external shock cooled electrons due to IC scatterings and that was
the dominant cooling mechanism for electrons radiating at radio/mm 
frequencies for $\sim10^2$ days. We note that the uncertainty in 
$\gamma_c$ is small as it depends primarily on $L_x$ which is well determined
from observations. However, $\gamma_i$ might be much larger than shown
in this figure, particularly for $p\approx2$ favored by our modeling of
the observed data, if the minimum Lorentz factor of electrons accelerated in
shocks is larger than the value we have adopted in this paper, viz. 
$\gamma_i = \epsilon_e (p-2)(\Gamma-1)m_p/[m_e(p-1)]$.
}
\label{fig2}
\end{figure}

Theoretically computed light curves for Sw J1644+57 in frequency bands
230 GHz \& 87 GHz, together with observed data from Berger
et al. (2012), are shown in Fig. 3. The theoretical light curves
were calculated from a decelerating blastwave which had a fixed
amount of total energy (i.e. no late time injection of energy) and yet 
the observed flux in these frequency bands is 
nearly flat for $\sim 10^2$days (consistent with the observed data)
in spite of the fact that these bands clearly lie above the 
synchrotron injection ($\nu_i$) and self-absorption ($\nu_a$) 
frequencies (shown in fig. 4). There is a noticeable dip
in light curves between $\sim10$ \& 20 days.
The reason for this dip is that the observed x-ray light curve during this
period is nearly flat (except, of course, for the ever present
fluctuations). A flat x-ray light curve means that $\nu_c$ does not change 
much during this time interval (see fig. 4). However, $\nu_i$ and the peak 
flux (which are independent of $L_x$) continue their monotonic decline, 
and that is the reason for the flux decrease during 
this period. A more realistic treatment of the x-ray flux
in our calculation, that includes the observed fluctuations, should provide
a better fit to the observed data.

\begin{figure}
\centerline{\hbox{\includegraphics[width=9cm, angle=0]{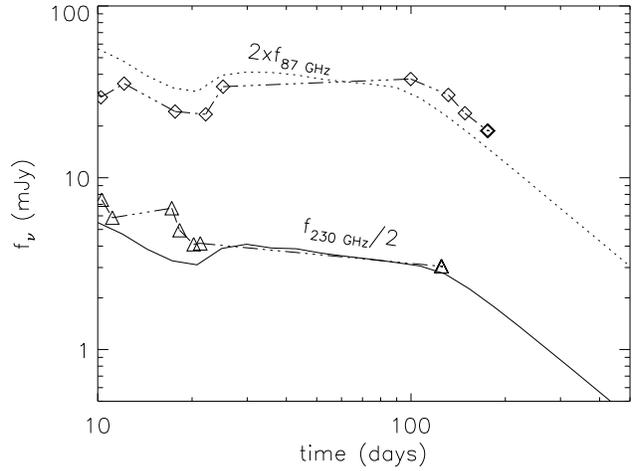}}}
\caption{Light curves in frequency bands 230 GHz (solid line) \& 87 GHz 
(dotted curve) calculated for the model described in \S3, together with 
observed data for Sw J1644+57 from Berger et al. (2012) are shown in this
figure. The parameters used for these calculations are same as that
in Fig. 2 (see fig. 2 caption for details). Light curve before 10 days
is not shown here because the wild fluctuation in x-ray flux at 
earlier times (by more than a factor 10) introduce huge error in our 
theoretical calculations which is based on a power law decline model
for x-ray luminosity for $t_{obs}>20$days, and a constant luminosity for 
$10{\rm d}\le t_{obs} \le 20$d. Light curves below 50 GHz are also not 
shown because of large errors in theoretical calculations at frequencies
less than $\sim 2\nu_a$ as discussed in \S4.
}
\label{fig3}
\end{figure}

The nearly flat light curves for $\sim 10^2$ days for 87 and 230 GHz is due 
to the fact that these bands lie above the synchrotron cooling frequency 
which rises very rapidly with time ($\nu_c\aprop t_{obs}^{3}$) due to 
the rapidly declining x-ray flux; we obtained this behavior analytically in 
the last section, and the numerical result is presented in fig. 4. 
The light curve remains nearly flat as long as the observing band is above
$\nu_c$, and then it starts to decline as $\sim t_{obs}^{-1.8}$. 
The duration of the flat phase is dictated by the temporal 
behavior of x-ray flux which we took for these calculations
to decline as $t_{obs}^{-5/3}$, and the spectrum was taken to be
a power law function; $f_\nu\propto\nu^{-0.7}$ for $0.1{\rm keV} \le \nu 
\le 10$keV as per Swift/XRT observations\footnote{Swift/XRT observes
photons with energy between 0.3 and 10 keV. However, it would be a remarkable
coincidence if the spectrum for Sw J1644+57 starts deviating from the
observed power law behavior exactly at the low or the high energy threshold
for XRT. The infrared k-band flux at 10 days was 0.03 mJy (Levan et al. 2011) 
and at the same time the mean 1 keV flux was 0.02 mJy (Zauderer et al. 2013),
and that suggests that the XRT spectrum became much harder, perhaps
$\nu^{1/3}$, at $\sim 0.1$ keV. The flux above 10 keV, however, is
rather poorly constrained by observations.}
for $t_{obs} < 40$ days, eg.  Saxton et al. (2012). 
The observed x-ray light curve in reality is much more complicated, and 
shows rapid fluctuations.
However, we do not expect $\nu_c$, and radio light curves, to follow 
these wild fluctuations because the signal we receive from
the external shock at any given fixed observer time is in fact an 
integral over equal arrival time 3-D hypersurface that has a large
radial width so that the effect of x-ray radiation on electron
distribution function is averaged over a time duration $\delta t_{obs}$ of 
order $t_{obs}$. The x-ray flux averaged over this time scale is not a smooth
power law function, but consists of periods of faster and slower declines.
The duration of time when light curve in an observed frequency
band is roughly flat depends on these 
fluctuations, and for that reason the duration of the flat phase 
shown in figure 3 should be considered to have an uncertainty of order
20\%.

The light curve decline for frequency $\nu$ which lies below the synchrotron 
cooling frequency ($\nu_c$) is easy to obtain using the scalings provided 
in the \S3.2.1. For sub-relativistic shocks, $B'\propto t_{obs}^{-3/(5-s)}$ 
(eq. \ref{Bp1}), and $\nu_i\propto t_{obs}^{-(15-4s)/(5-s)}$ 
(eq. \ref{nui}), and so the observed flux for $\nu_c > \nu > 
\max(\nu_a, \nu_i)$ decreases with time as $t_{obs}^{-[(15-4s)(p-1) - 
(6-4s)]/2(5-s)}$; for J1644+57, $\nu_c\sim100$ GHz at $t_{obs}\sim10^2$d 
(Fig. 4). For $s=2$ and $p=2.3$, the light curve decline is expected to be 
$t_{obs}^{-1.85}$ which is consistent with the observed decline 
(see the 87 GHz light curve in Fig. 3 on this page, and also 8.4 --- 43 GHz 
light curves in Fig. 3 of Zauderer et al. 2013).

The reason that we do not show theoretical light curves in fig. 3 for 
$t_{obs}<10$ days is that the x-ray
light curve before 10 days is dominated by very large flares ---
the flux decreased by a factor $\sim10^2$ between 4 and 6 days
--- which cannot be captured by a simple power law
model used in this paper. A calculation
of IC cooling of electrons, and the emergent flux, before 10 days must
include these wild fluctuations for the result to be meaningful.

\begin{figure}
\centerline{\hbox{\includegraphics[width=8.25cm, angle=0]{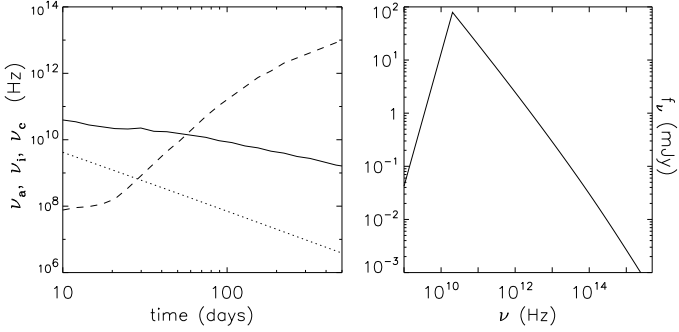}}}
\caption{The left panel shows the three synchrotron characteristic
frequencies, obtained by numerical calculations, for the same set of 
parameters as fig. 2; see fig. 2 caption for details; solid line is
$\nu_a$, dotted line is $\nu_i$ and the dashed curve is $\nu_c$. Note 
the sharp rise in $\nu_c$, as $\sim t_{obs}^3$, between 20 and 100 days, 
which is consistent with our analytical estimates (see \S3).
We note that the uncertainty in the calculation of $\nu_c$ is small. However, 
the actual value of $\nu_i$ might be larger than in this figure by 1--2 orders 
of magnitude for $p\approx2$ favored by our modeling of 
the observed data (see caption for Fig. 2 for explanation); good fits to
observed light curves (like those shown in fig. 3) can be obtained even when
$\nu_i$ is much larger than shown in this figure by appropriately rescaling 
microphysics parameters and changing $E$ and $n$ by a factor $\sim2$. 
The right side panel shows the spectrum calculated at 52 days. The spectrum
above $\nu_i$ and $\nu_c$ is $\nu^{-0.8}$ even though we took
$p=2.3$. This is due to the fact that when cooling is dominated by IC 
scatterings of x-ray photons in the K-N regime, the spectrum is 
$\nu^{-(p-1+\beta)/2}$ instead of $\nu^{-p/2}$; the spectral index for x-ray 
radiation was $\beta=0.4$ for $t_{obs}\gae 50$d (Saxton et al. 2012). The 
spectrum steepens to $\sim\nu^{-1.15}$ for $\nu\gae10^{13}$Hz when the 
cooling is dominated by synchrotron loss.
}
\label{fig4}
\end{figure}

The rise in light curves at lower frequencies (20 GHz or less) and 
for $10{\rm d}<t_{obs}\lae10^2$d is due to $\nu_a$ lying above the band
(see Fig. 4). This behavior can be understood using the analytical scalings 
presented in the last section. However, the error in the calculation of 
emergent light curves for frequencies $\lae 2\nu_a\sim 50$GHz is very 
large if the shell of shocked CNM is treated as homogeneous as we have 
done in this work. The reason is that 
fluctuations in $L_x(t_{obs})$ lead to variation of
synchrotron absorptivity with angular position and distance from the 
shock front (due to the effect of X-rays on electron distribution function).
And due to the exponential dependence of the emergent flux on the synchrotron 
optical depth along the photon trajectory, the error in theoretically
calculated light curves is large when these variations are ignored. At 
sufficiently high frequencies, $\nu \gae 50$GHz, the optical depth is 
small and spatial variations of synchrotron absorption are
averaged out, and therefore, treating the source as homogeneous is 
a fine approximation for the calculation of high frequency light curves. 
This is the reason we show light curves above about 
50 GHz in Fig. 3, and not at lower frequencies.

The solutions we are finding have $\min(\nu_i, \nu_c)<\nu_a$, and that might
naively suggest that the spectrum below $\nu_a$ should be $\nu^{2.5}$ in
conflict with the observed spectrum of $\sim\nu^{1.5}$ for $t_{obs}\lae 10^2$d.
The fact that the observed spectrum is shallower than $\nu^2$ is already telling
us that something is missing. And most likely what that is, is that the jet 
has some angular structure, and when integration over equal 
arrival time hyper-surface is carried out to obtain the observed spectrum,
it turns out to be softer by $\nu^{0.5}$---$\nu^1$ than the comoving, 
instantaneous, spectrum\footnote{The equal arrival time hyper-surface is 
a 3-D section of the 4-D space--time with the property that radiation from 
every point on this hyper-surface arrives at the observer at the same time. 
The radius ($R$), $\gamma_i$, $B'$ etc. for different points on the 
hypersurface are in general different for a non-uniform jet, i.e. a jet 
where energy density, Lorentz factor etc. vary with angle wrt jet axis
and distance from the jet surface. It can be shown that an integration
over equal arrival time hyper-surface invariably softens the observed
spectrum.}.

The infrared flux at 10 days (0.03 mJy, Levan et al. 2011) is roughly
of order the theoretically calculated value for parameters used for fitting
230 GHz light curve shown in Fig. 3.  The infrared producing electrons have 
$\gamma_e' \sim 10^4$ and their IC cooling by scattering x-rays photons is 
highly suppressed due to K-N effect. However, $\nu_c$ due to synchrotron 
cooling is $\sim 10^{13}$Hz, so the spectrum in this band should be 
$\sim \nu^{-1}$ if the emission is dominated by the external shock. 
Levan et al. (2011) report that the infrared light curve 
declined with time as $\sim t_{obs}^{-3}$ between 5 and 10 days. 

This is much too steep to be explained by external shock emission, and we 
suspect that this fall off suggests that IR radiation is produced by a 
process internal to the jet. The claim of 7.4$\pm$3.5 \% linear 
polarization in the IR K$_s$-band, at $\sim17$days after the Swift/BAT
trigger, by Wiersema et al. (2012) supports this scenario.

A theoretically calculated spectrum for external shock synchrotron
radiation at 52 days is shown in Fig. 4. The spectrum above the peak 
is quite hard, $\sim \nu^{-0.7}$, even though $\nu_i$ and $\nu_c$ lie 
below the peak. The reason is that when electrons are cooled by IC
scatterings in K-N regime the resulting spectrum is
$\nu^{-(p-1+\beta)/2}$ (see \S3.1) and not $\nu^{-p/2}$. X-ray data show that 
$\beta$ varies with time (Burrows et al. 2011), and that $\beta$
was about 0.7 for $t_{obs}\lae 50$d and $\sim 0.4$ for $t\gae 50$d
(Saxton et al. 2012). Therefore, we expect the radio spectrum above the
peak to be $\sim\nu^{-0.85}$ at early times, and $\nu^{-0.7}$ at late times,
for $p=2$. The late time radio data are consistent with this expectation, 
and for $t_{obs}\lae 50$d spectra above the peak are poorly sampled.

\subsection{High energy radiation}
\label{LAT-limit}

IC scattering of an x-ray photon of frequency $\nu_x$ by an electron of 
Lorentz factor $\gamma_i'$ in the external shock produces a
high energy photon of energy $\sim 2\gamma_i'^2 \nu_x$. Considering that 
$\gamma_i'\sim30$ (Fig. 2), the IC photon energy is $\sim 10$ MeV, and the
expected IC luminosity is $\sim\tau_T L_x \gamma_i'\min(\gamma_c',\gamma_i')$; 
where $\tau_T \sim \sigma_T n(R) R\sim 10^{-5} R_{17} n_{f,2}$ is 
Thomson optical depth. At $t_{obs}=10$d, $L_x\sim 3\times10^{46}$erg s$^{-1}$,
$\gamma_c'\sim4 R_{17}$ \& $\gamma_i'\sim 30$ (fig. 2), and therefore we expect 
IC luminosity to be $\sim 5\times10^{43}R_{17}^2 n_{f,2}$ erg s$^{-1}$.
Another way to estimate the luminosity of high energy IC photons
follows directly from the fact that IC scatterings of x-ray photons is
efficient at cooling electrons at early times, i.e. IC cooling time is 
less than or of order the dynamical time for $t_{obs} < 20$d. Hence, the 
IC luminosity should be roughly equal to the energy luminosity carried by 
electrons accelerated by the external shock, i.e. $L_{ic} \le 16\pi 
\epsilon_e R^2 m_p c^3 n(R) \Gamma^2\sim 2\times10^{44} \epsilon_{e,-1} 
n_{f,2} R_{17}^2$erg s$^{-1}$. There is no data between 1 \& 100 MeV 
for Sw J1644+57 to our knowledge,
however, Fermi/LAT provided an upper limit of about 10$^{46}$erg/s 
above 100 MeV at 10d (Burrows et al. 2011). This suggests that 
$n_{f}$, the CNM density at $R=10^{17}$cm, cannot be larger than
about 10$^4$cm$^{-3}$.

\subsection{Parameters of the TDE and the external shock}

An accurate determination of jet energy and other parameters for Sw J1644+57
should take into consideration the effect of fluctuating $L_x$ on electron 
distribution, a more detailed treatment of the dynamics than described by 
equation (\ref{E_conserve}), and the 3D structure of shocked plasma 
and magnetic fields. But that is beyond the scope of this work\footnote{The 
goal of this work is to show that cooling of electrons by x-ray photons 
passing through the external shock can explain the puzzling, flat, light curves 
observed in mm/cm bands for $\sim10^2$days.}. We show in Fig. 5 an
approximate determination of $E$ (isotropic energy in the external shock), 
the CNM density, and $\epsilon_B$ (energy fraction in magnetic fields) allowed 
by the multiwavelength data for Sw J1644+57; see the caption for Fig. 5
for a description of the various observational constraints that were
used for the determination of these parameters.

Fermi/LAT upper limit restricts the CNM density at 10$^{17}$cm to be smaller 
than $\sim5{\rm x}10^3$cm$^{-3}$ (see \S\ref{LAT-limit}). And according to the 
radio/mm/infrared data the isotropic energy in the blast wave
is $\sim 2\times10^{53}$erg which is consistent with estimates obtained
using equipartition argument, e.g. Barniol Duran et al. (2013). We note 
that the uncertainty in these parameters and $\epsilon_B$ is a factor of 
a few. It should be 
pointed out that the afterglow data we have used for determining energy etc.
consists of observations carried out 10 or more days after the TDE.
The blast wave Lorentz factor at this time had dropped to $\lae 2$, and the 
angular size of the shocked CNM is perhaps larger by a factor
$\sim 2$ than the jet angle at the smaller radius where x-rays were produced.
Thus, the beam corrected energy in the external shock is $\sim 10^{52}$erg.
A more detailed discussion of beam corrected energy is provided at the
end of this sub-section.

We have also looked for the density stratification ($s$) and find that
$s\gae 1.5$ provides a reasonable fit to the available data; $s=1.5$
density stratification is expected for an advection dominated accretion 
flow in the neighborhood of a black hole, and $s=2$ is when outflow
determines the CNM density structure.

Although we took $\epsilon_e=0.1$ for results shown in all of the figures, 
other values of $\epsilon_e$ are allowed by the data as long as we 
take $E\aprop\epsilon_e^{-1}$.

The 1 keV x-ray flux at 610 days after Swift/BAT trigger --- for 
the entire solution space shown in Fig. 5 --- is within a factor 
$\sim2$ of the value observed by the Chandra satellite which was 0.7nJy
(Levan \& Tanvir, 2012).

\begin{figure}
\centerline{\hbox{\includegraphics[width=8.25cm, angle=0]{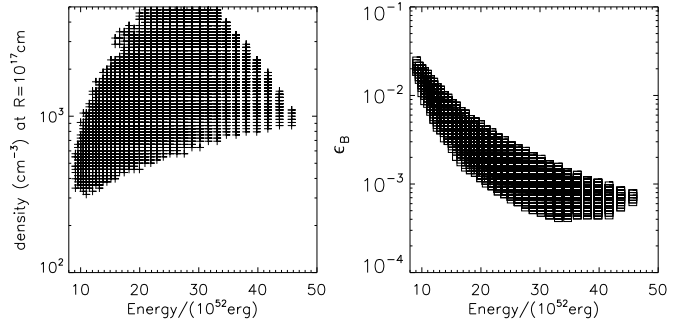}}}
\caption{Isotropic blast wave energy ($E$), CNM density at $R=10^{17}$ cm,
and $\epsilon_B$ allowed by the data for Sw J1644+57. The parameters shown
in this figure should be considered a rough estimate due to the approximate 
way we have handled the blast wave dynamics, and used a smooth decline for the 
x-ray flux passing through the external shock and cooling electrons 
instead of the highly fluctuating flux Swift/XRT observed. The observational
constrains that were used in the determination of parameters are: (1) flux at 
230 GHz at 10 days after the tidal disruption was 10 mJy (Berger et al. 2012), 
(2) $\nu_a=15$ GHz at 10 days, (3) the flux at the peak of the spectrum at 
10$^2$days is 50 mJy, (4) an upper limit of 10$^{46}$erg s$^{-1}$ for 
$>$10$^2$MeV at 10 days which suggests that $n_f\lae 10^4$cm$^{-3}$ 
($n_f$ is density at $R=10^{17}$cm); a tolerance of $\pm30$\% was used for 
all these constraints which reflects the degree of uncertainty of our 
calculations. We took $L_x = 4\times10^{46}{\rm erg/s} (t_{obs}/20 
{\rm days})^{-5/3}$ for $t_{obs}>20$d and constant between 10 and 20 days,
and $\epsilon_e=0.1$ for these calculations; for $\epsilon_e=0.5$ 
these plots look very similar, except that they are shifted to the left 
by a factor of $\sim 5$. The true energy in the blast wave, corrected for its 
finite angular size, is a factor of $\sim10$ smaller than the isotropic
energy shown in the figure (see the last part of \S4.2 for a detailed 
discussion).}
\label{fig5}
\end{figure}

Figure 5 shows isotropic equivalent of energy in the blast wave. The
true, beam corrected, energy in the explosion can be obtained from
this using the blast wave Lorentz factor, $\Gamma_j$, and the angular size of the beam.
The transformation from an isotropic to a finite opening angle source
preserves the specific flux in observer frame which is given
here in terms of specific intensity in the source comoving frame ($I'_{\nu'}$)
\begin{equation}
   f_\nu = {2\pi\over (1+z)^3 \Gamma_j^3} {R^2\over d_A^2} \int_0^{\theta_j}
   d\theta {\sin\theta\cos\theta I'_{\nu'}\over (1 - v_j\cos\theta)^3}, 
\end{equation}
where $\theta_j$ is half-opening angle of the source, $z$ is its redshift,
$d_A$ is angular diameter distance, and $v_j$ is the ratio of shock front speed
and speed of light. Treating the angular variation of 
the specific intensity in the jet comoving frame to be small we find
\begin{equation}
   f_\nu = {\pi I'_{\nu'} R^2\over (1+z)^3 \Gamma_j^3 d_A^2} \left[
   {2v_j -1\over v_j^2(1-v_j)^2} - {2v_j\cos\theta_j-1\over v_j^2
   (1-v_j\cos\theta_j)^2} \right].
\end{equation}
The requirement that blast waves of different opening angles
produce the same observed flux ($f_\nu$) requires appropriate 
rescaling of $I'_{\nu'}$.
Thus, it follows from the above equation that the energy in a blast wave
with half opening-angle $\theta_j$, which is proportional
to $(1-\cos\theta_j) \int d\nu' I'_{\nu'}$, is smaller than a spherical blast wave of
the same LF by a factor:
\begin{equation}
    {E_{\theta_j}\over E_{iso}} = {(1-\cos\theta_j)\over (1 - v_j)^2 }
   \left[ {2v_j - 1\over v_j^2 (1-v_j)^2 } - {2v_j\cos\theta_j-1 \over
    v_j^2 (1 - v_j\cos\theta_j)^2 } \right]^{-1}.
\end{equation}

An upper limit to angular size $\theta_j$ can be obtained from the 
maximum transverse speed of spreading in the source comoving frame
which is equal to the sound speed of the shock heated plasma given by
\begin{equation}
  c_s = v_s \left[ {2\hat\gamma(\hat\gamma-1) \over (\hat\gamma + 1)^2}
   \right]^{1/2},
\end{equation}
where $\hat\gamma$ is the ratio of specific heats for shocked plasma which is
between 4/3 and 5/3 due to the fact that electrons are relativistic
whereas protons are sub-relativistic at late times for the external shock
of J1644+57, and $v_s$ is the speed of the unshocked fluid wrt
the shock front ($v_s > v$; $v$ is the speed of the shocked fluid wrt
unshocked CSM).
The perpendicular size of the system, $R_{\perp}=R\theta_j$, varies as
$dR_{\perp}/dR=c_s/(v_s \Gamma_s)$, if the jet spreads sideways at the speed $c_s$.
Therefore, the angular size of the blast wave when it is at radius $R$ is:
 $\theta_j < [\theta_0 R_0 + (R-R_0)c_s/(v_s\Gamma_s)]/ R \sim
 (c_s/v_s\Gamma_s) (1  - R_0/R)$; where $\theta_0$ is the angular size when
the blast wave was at radius $R_0$. Considering that $c_s/v_s \sim 0.5$,
that $\Gamma_s\gae1.3$ (since $v_s>v$), and that
$R$ increases by a factor of $\sim10$ between 10 and 500 days, the jet 
angle at $\sim 500$ days cannot be larger than
$\sim 0.3$ radian; most likely $\theta_j\sim 0.2$ at 500 days
since the sideways expansion speed becomes $c_s$ only when the shocked plasma
cools down to a temperature substantially lower than the temperature it
had just behind the shock-front\footnote{Kumar \& Granot (2003)
using 1-D hydro simulations found little evidence for sideways spreading
for a beamed jet interacting with ISM during the relativistic phase,
and the 2-D simulation of Zhang \& MacFadyen (2009) found the 
jet angle at 500d -- when the blast wave was sub-relativistic -- to be
close to 0.2 radian which should be independent of the choice of the 
initial jet angle.}.

For $\Gamma_j = 1.2$ and $\theta_j=0.2$, $E_{\theta_j}$ is smaller
than an equivalent spherical blast wave by a factor 4.3, and for $\Gamma_j=1.5$,
and $\theta_j=0.2$, $E_{\theta_j}$ is smaller by a factor $\sim7$.
The isotropic energy we show in Fig. 5 is for $\epsilon_e=0.1$. The
energy in the external shock is smaller by a factor $\sim 10\epsilon_e$
for a larger $\epsilon_e$. So the isotropic energy shown in Fig. 5 should
be divided by a factor $\sim10$ (if $\epsilon_e = 0.2$, which is very likely)
in order to obtain the true energy in J1644+57 as required by the radio/mm data.
Therefore, the total energy in the blast wave of J1644+57 is estimated to be 
of order $10^{52}$ erg which is consistent with the amount obtained 
independently from the x-ray data.

\subsection{A puzzling drop in $\nu_a$ between 5 \& 10 days}

According to Berger et al. (2012) the synchrotron-absorption frequency ($\nu_a$)
dropped by a factor 10 between 5 and 10 days after the tidal disruption
event Sw J1655+57. If these observations are correct\footnote{We would not be
surprised if it were to turn out that the drop in $\nu_a$ has been 
overestimated by observers by a factor 2--3. The spectrum reported by 
Berger et al. (2012) at $t_{obs}=10$d has no frequency coverage between 
15 and 80 GHz. It is possible that the actual peak was at $\sim 50$GHz, 
in which case $\nu_a$ at 10d has been underestimated by a factor $\sim 3$.} 
then this result essentially rules out any shock model, with or without 
continuous energy injection, for radio emission unless some mechanism 
external to the shock led to a drastic change to the electron energy and 
magnetic fields in shocked plasma between 5 and 10 days. To understand 
this result we make use of equation (\ref{nua1}) for $\nu_a'$:
\begin{equation}
      \nu_a'^2 \gamma_e'(\nu_a') \propto B' N_e(\ge\gamma_e') \;{\rm or}\;
      \nu_a \propto \Gamma \left[ B'^{3/2} N_e(\ge\gamma_e') \right]^{0.4}
   \label{nua4}
\end{equation}
where $\gamma_e'(\nu_a')$ is the Lorentz factor of electrons that have 
characteristic synchrotron frequency $\nu_a'$, and $N_e(\ge\gamma_e')$ 
is the column density of electrons that radiate above $\nu_a'$. The right 
hand side of the first equation is proportional to the synchrotron flux 
at $\nu_a'$ in the comoving frame of the source when synchrotron absorption 
is ignored, and in the second equation we have considered the case where 
$\nu_a'>\min(\nu_i', \nu_c')$ (other cases are straightforward to consider).
For any external shock model, with or without energy injection and arbitrary
CNM density stratification, $\Gamma$, $B'$ and $N_e$ are smooth functions 
of time (see \S3). Since $\Gamma \propto t_{obs}^{-(3-s)/(8-2s)}$ (eq. 
\ref{dyna1}), the most change we can expect for $\Gamma$
between 5 and 10 days is a factor $2^{3/8}\sim1.3$ for a $s=0$ medium;
similarly, $B'\propto \Gamma R^{-s/2}$ cannot change much during this time
in the absence of an external agency at work. And $N_e$, which depends on
the number of electrons in the CNM swept up by the shock, is typically an
increasing function of time and hence its effect on $\nu_a'$ is to increase it
\footnote{If a shock were to run into a dense cloud, that would cause
$\nu_a$ to increase. And when a shock suddenly encounters a zero density medium,
the shocked plasma undergoes adiabatic expansion and then $\gamma_e'$,
$B'$, and $N_e'(\ge\nu_a')$ decrease with time. It is easy to show that the 
net effect of adiabatic expansion is a slight decrease of $\nu_a$ in time 
interval $t_{obs}$---$2 t_{obs}$.}. We know empirically that $N_e(\ge\gamma_e') 
B'\propto f(\nu_a)/\Gamma$ decreased 
by a factor $\sim3$ between 5 and 10 days for Sw J1644+57 (see fig. 2 of Berger 
et al. 2012), and that by itself can only account for a decrease of 
$\nu_a$ by a factor $\lae 2$. We see, therefore, that unless there is 
a dramatic change in $B'$ and/or $\gamma_e'(\nu_a')$ on a short time scale 
of $\delta t_{obs}/t_{obs} = 1$ --- which is not possible in external 
shocks without a powerful radiation front passing though it or some such 
mechanism --- $\nu_a$ cannot drop by a factor $\sim10$ between 5 and 10 days 
that is observed for Sw J1644+57; any addition of energy to a decelerating 
shock only makes the problem worse.

It is probably no coincidence that the x-ray luminosity of Sw J1644+57 fell by 
a factor $\sim50$ during the same time period that $\nu_a$ dropped by a factor 
10. When a highly luminous x-ray front was passing through the external shock 
($t_{obs}\lae 6$d), electrons in the region were cooled by IC scatterings of 
x-rays to a Lorentz factor of order unity in less than one dynamical time
(see eq. \ref{tic_tdy}). The sharp drop in x-ray luminosity during 6 to 8 days
by a factor 50 reduced the IC cooling of electrons and raised $\gamma_c'$ by  
the same factor. As we can see from equation (\ref{nua4}) a rising
electron temperature helps to lower $\nu_a$, however, the effect on 
$\gamma'(\nu_a')$ is much smaller. Ultimately, a sharp drop in $B'$ is 
needed to account for the observed decrease of $\nu_a$ as can be
seen from the second part of equation (\ref{nua4}). 
 We suspect that a sharp drop in $L_x$ leads to a smaller $B'$.
One possible way this might occur is that while $L_x$ is large, or jet power 
is high, the interface separating the shocked jet and the shocked CNM is 
Rayleigh-Taylor unstable and the turbulence associated with this instability 
produces strong magnetic fields. Currents driven in the shocked CNM by high 
$L_x$ might be another way to generate close to equipartition magnetic 
field in the region where radio/mm emission is produced. However, a connection
between $L_x$ and magnetic field generation is just a speculation at this 
point\footnote{We do not expect the magnetic field to continue to decline 
with $L_x$ indefinitely. The reason for
this is that below a certain threshold x-ray luminosity magnetic field
generation in external forward shock proceeds by Weibel mechanism, or some other
instability, that is not affected by $L_x$.}. 

\section{Concluding remarks}

An interesting puzzle is posed by the radio/mm data for the tidal disruption 
event Swift J1644+57: light curves in frequency bands that lie above the 
synchrotron characteristic peak frequencies, i.e. $\nu>10$ GHz, are roughly 
flat for $10^2$days rather than decline with time as $\sim t_{obs}^{-1}$ 
as one would expect for radiation produced in the external shock. 
Berger et al. (2012) suggested that the flat light curves are due to 
continued addition of energy to the external shock for 10$^2$ days.
This requires about 10 times more energy to be added at late times 
than is seen in x-rays, and that is worrisome. We have described in 
this paper a simple model for flat light curves that does not require
adding any extra energy to the decelerating shock wave. The flat light curves, 
instead, are a result of efficient cooling of electrons in the external shock
by IC scatterings of x-ray photons that pass through the external shock region 
on their way to us (Fig. 1). The IC cooling becomes weaker with time as the 
x-ray flux declines and that is the origin of radio/mm light curve flatness. 
This model can account for the radio and milli-meter data (temporal decay 
as well as spectra), and is consistent with $>$10$^2$MeV upper limit 
provided by Fermi/LAT.

The angular size of the x-ray beam should not be much smaller than 
the size of the external shock region during the initial $\sim10^2$days
for the IC cooling model to work; otherwise only a fraction of shock 
accelerated electrons will be exposed to IC cooling, and the effect 
described here is weakened. The angular spreading of blast wave is
discussed in \S4.2 where it is shown that even at $t=500$d the 
angular size of the jet is no more than a factor 2 larger than its
size at 10d. Moreover, the decline of the late time radio/mm light curves 
as $\sim t_{obs}^{-1.8}$ for $t_{obs}\gae 200$days 
(Zauderer et al. 2013; Fig. 3) requires --- in the framework of
synchrotron radiation from a decelerating blast wave model --- 
the angular spreading to be very weak; the light curve decline for
a sub-relativistic blast wave of a constant opening angle, for $p=2.3$
\& $s=2$, can be easily shown to be $t_{obs}^{-1.85}$ using formulae
provided in \S3.2.
In fact, considering the {\it naturalness} of this 
model --- it is unavoidable that x-ray photons pass through the external 
shock and cool electrons --- and its clear cut, robust, implications for
external shock light curves, it is not unreasonable to conclude from the
data for this TDE that sideways spreading speed is at best modest for 
sub-relativistic blast waves as long as $\Gamma\gae1.1$. 

The model predicts a connection between the observed x-ray light curve and 
the temporal and spectral properties of radio and mm radiation from 
Sw J1644+57. It requires more extensive numerical modeling than presented 
in this paper to explore this connection in detail, which can then be
confronted with the extensive data for this remarkable event. 

Barniol Duran \& Piran (2013) have suggested two broad class of models
for Sw J1644+57 using equipartition arguments within the context of 
synchrotron emission, but without considering the IC cooling of electrons. 
The first scenario consists of one jet that is responsible for both 
the X-ray and the radio emission. The second scenario invokes two sources: 
a narrow relativistic jet to produce X-rays, and a quasi-spherical,
mildly relativistic, source to produce the radio emission. Both models
require the energy available to electrons for synchrotron radiation to 
increase with time, but Barniol Duran \& Piran (2013) argue that the 
total energy in the radio source could be constant. The model suggested 
in the present work roughly corresponds to the first scenario, with a 
constant total energy, where the energy fraction available to electrons 
for synchrotron radiation increases with time as the x-ray luminosity 
decreases and the IC losses weaken.

\section*{Acknowledgements}

We thank Patrick Crumley and Roberto Hernandez for comments on the manuscript.

\end{document}